\RequirePackage{fix-cm}
\documentclass[twocolumn]{svjour3}          

\usepackage[T1]{fontenc}
\usepackage[utf8]{inputenc}
\usepackage{newtxtext,newtxmath}
\usepackage{graphicx}
\usepackage{epstopdf}
\usepackage{epsfig,graphics}
\usepackage{latexsym}
\usepackage{multirow}
\usepackage{booktabs}
\usepackage{cite}
\usepackage{citeref}
\usepackage{subfigure}
\usepackage[misc]{ifsym}
\usepackage[export]{adjustbox}
\usepackage[T1]{fontenc}
\usepackage{hyperref}
\usepackage{mathtools,xparse}
\usepackage{pgfplots}
\usepackage{tikz}
\usepackage{tkz-euclide}
\usetkzobj{all}
\usetikzlibrary{positioning}
\usetikzlibrary{shapes,arrows}
\usetikzlibrary{shadows, positioning,arrows.meta,shapes.geometric}
\usepackage[flushleft]{threeparttable}
\usepackage{floatrow}
\usepackage{wrapfig}
\usepackage{algorithm}
\usepackage[noend]{algpseudocode}
\usepackage{etex}
\usepackage{lipsum}

\begin{document}

\title{Deep Learning with ConvNET Predicts Imagery Tasks Through EEG}




\author{Apdullah Yay{\i}k \and Yakup Kutlu \and G\"{o}khan Altan}



\institute{A. Yay{\i}k (\Letter)  \at Military Academy \\National Defence University, Ankara, Turkey\\  email: ayayik@kho.edu.tr\\
Y. Kutlu \at Department of Computer Engineering,\\\.{I}skenderun Technical University, Hatay, Turkey\\
G. Altan \at Department of Computer Engineering,\\\.{I}skenderun Technical University, Hatay, Turkey
}


\maketitle
\begin{abstract}
Deep learning with convolutional neural networks (ConvNets) have dramatically improved learning capabilities of computer vision applications just through considering raw data without any prior feature extraction. Nowadays, there is rising curiosity in interpreting and analyzing electroencephalography (EEG) dynamics with ConvNets. Our study focused on ConvNets of different structures, constructed for predicting imagined left and right movements on a subject-independent basis through raw EEG data. Results showed that recently advanced methods in machine learning field, i.e. adaptive moments and batch normalization together with dropout strategy, improved ConvNets predicting ability, outperforming that of conventional fully-connected neural networks with widely-used spectral features.

\end{abstract}

\section{Introduction}
\label{intro}
Machine learning methods together with electroencephalography (EEG) data empowers researchers to interpret neurological activities, and are key components of brain-computer interface (BCI) research field. For instance, such systems can enable locked-in patients to type phone numbers \cite{ref1}, to use wheel-chair \cite{ref2} and to operate computer explorer \cite{ref3}. In addition, such systems may be used in prediction onset of epilepsy \cite{ref4}. Although these successful and promising studies, a general framework for extracting features and learning mechanism with regard to recent advances in machine learning field is still needed.

Deep learning with convolutional neural networks (ConvNets) are of prominent recent advances in machine learning, particularly computer vision. They are the most successfully biologically inspired neural networks since their principles and structures rely on nonscientific hierarchical learning \cite{ref5}. Following achieving great success in computer vision, it continued in
a straight way in sentiment analysis from text \cite{ref6} and audio processing \cite{ref7}. Nowadays, handcrafted-features have lost their usefulness with ConvNets capability to reveal prominent features from input data via end-to-end hierarchical representation. 

This paper concentrated on a challenging task of predicting imagined left and right movements through raw EEG data with ConvNet on a subject-independent basis with considering 109 number of subjects. In the literature, studies on EEG motor movement/imagery (EEGMMI) database aim to predict imagined movements through use of support vector machine or multi-layer perceptron (MLP) achieved success on either only a subject-dependent basis or a subject-independent basis but for limited subjects. Mostly, it is claimed that this specialized tasks could uniquely predicted just for each individual subject. Additional researches were performed in distinguishing executed and imaginary motor movements \cite{ref12}, \cite{ref13} that differ form our study in that we concern predicting imagined motor left and right fist movements. Mohammed et al. proposed SVM learning model for predicting motor imagery activities based on wavelet spectral analysis. They have reached accuracy of 84\% on a subject independent basis for only 20 subjects \cite{ref14}. 

Besides, studies not-using EEGMMI database reached promising results with considering artifact removal at preprocessing and energy and power features \cite{ref15}, proposing Joint Approximate Diagonalization method for handling non stationary characteristics of EEG that aids in predicting imagined movements \cite{ref16}, integrating magnetoencephalographic (MEG) signals with EEG and converting EEG time-series into 2D mesh-like hierarchy together with convolutional recurrent neural network \cite{ref17}. 

EEG data are physically dissimilar to typical 2-D or 3-D images input of ConvNets, they consists of time-series from several electrodes on scalp surface, can be conceptualized as 2-D, voltage varies over time and space, where space refers to electrodes. In neuroscience field, EEG data are assumed to be originated from several dipolar current sources in the brain and they are linear combinations of these. From this perspective, spatial relations should be preserved and are of key components in EEG data to reveal data of high signal-to-noise-ratio from that of low signal-to-noise-ratio. Therefore, adaptation of ConvNets inputs for EEG data should be handled. In addition, design-choices and learning strategies should be compared. 

In our study, a design-choice that preserves spatial information of multi-channel EEG data, includes dropout layer \cite{ref8} and batch-normalization \cite{ref9} and with different back-propagation methods i.e. RmsPROP \cite{ref10}, Adam \cite{ref11} and stochastic gradient descent with momentum were evaluated with the same hyper-parameters values i.e. learning rates, regularization constants.
To see impacts of ConvNet model on EEG data results of classical spectral features together with traditional fully-connected multilayer perceptron were compared. Results showed that recently advanced methods in machine learning field, i.e. Adam, batch normalization together with dropout strategy, improved predicting ability, outperformed that of conventional fully-connected neural networks with spectral features estimated with Welch periodogram.

\section{Materials and Methods} 
\label{sec:Materials and Methods}
First, information about EEG recordings used in this study and preprocessing were provided. This followed by describing Welch and Morlet wavelet methods of spectral analysis. Next, we explained ConvNet constructed for this study in detail, particularly the design-choice for EEG data. Afterwards, three training strategies were described.

\subsection{Database Description}
We evaluated predicting imagery left and right movements on publicly available EEGMMI dataset \cite{ref18} in Physionet \cite{ref19}. Dataset consists of 160 Hz sampled EEG recordings through 64 electrodes from 109 subjects in the course of 4 motor/imaginary tasks. Each subject performed 14 experimental runs: two one-minute baseline runs (one with eyes open, one with eyes closed), and three two-minute runs of each of the 4 following tasks.

In this study EEG recordings in the course of one of the tasks were considered. The procedure in the selected task is as follows: A target appears on either the left or right side of the screen, the subject imagines opening and closing the corresponding fist until the target disappears. Then the subject relaxes. This trial is repeated 3 times, each repetition has 15 number of right and left labeled segments. Therefore, for each subject there exist 45 number of labeled segments.

\subsection{Preprocessing}
Preprocessing was performed at a minimum level to enable ConvNet to capture dynamics and characteristics of EEG recordings itself without bias. EEG recordings were filtered above 30 Hz using designed high-pass filter with ordinary $3^{rd}$ order Butterworh filter.

\subsection{Multi-Layer Perceptron}
In this study, the network contained two fully-connected hidden layer comprising 100 and 75 nodes, respectively. The training set was segmented in estimation and validation subsets (85 and 15\% of the training set respectively). The tangent hyperbolic activation function was used for the hidden layers and output layer. The sequential (in other words, batch size is one) learning strategy was performed for computing gradients. Gradients were computed with steepest descent algorithm and a learning rate of 0.01 was set and kept constant throughout the training process. The training of the network was stopped either at the $100^{th}$ epoch or whenever the updates of the weights failed to reduce the loss (mean sum squared error) of the validation subset for 15 consecutive times. The status of the neural network was then reverted to the last most successful epoch.

\subsection{Welch Method}
Welch method includes dividing time series data into overlapped segments, estimating periodogram of windowed each segments using fast fourier transform and averaging \cite{ref20}. Dividing trials into overlapped segments provides more accurate estimation from non-stationary time series. However using same repetitive information cause problems in spectral analysis. To eliminate such repetitive information due to overlapping segments, non-rectangular windowing methods are used. By this way, amplitude of the data is attenuated at initial and last parts of segments therefore their unnecessary (repetitive) information are decreased. Of several windowing methods, Hann tapering is mostly preferred because it makes initial and last parts of segments fully equal to zero \cite{ref21}. Also, averaging enables to
estimate periodograms that have relatively lower variance than entire time series.

Each trial, which had a duration of 0.4 second (656 number of data) was split in Hann windowed segments of 0.15 ms length that overlaps 50\% with the previous segment $-$except for the first one and periodograms were estimated with resolution of 1.67. The estimated periodograms of alpha bands (8$-$12 Hz) with Welch method were used as features to train multi-layer perceptron.

\subsection{Deep Convolutional Neural Network}
Deep learning with ConvNets \cite{ref22}, \cite{ref23} are of specialized type of neural networks that particularly process grid-like shaped data. They have strong ability to learn non-linearly separated features by means of discrete convolutions and non-linear activations. In addition employing deep (multiple) layers allow them to represent high-level features as combination of low-level features. For affine transformation, they simply use widely-known discrete convolution operation in at least one of their layer rather than general matrix multiplication. Discrete convolution with weight-sharing enables convolutional layers to be efficient in representation of scale large scale of data (images, audio, etc) and equivariance to translation (that means shifting of input can easily be captured by naturally shifting discrete convolution). Following, elementwise non-linear activation functions i.e. ReLU, LeakyReLU are applied to improve separability of data. Pooling layer is typically applied following convolution layer that compresses (in a way of down-sampling) output groups of discrete convolutions in-line. Changing level of striding in convolutional layer also provides such compression. Pooling operations are generally performed with a function of L2 norm, maximum, mean or weighted mean. Such pooling operations make outputs gain almost invariant to tiny translations of the network input.

\begin{figure}
    \centering
    \includegraphics[scale=0.75]{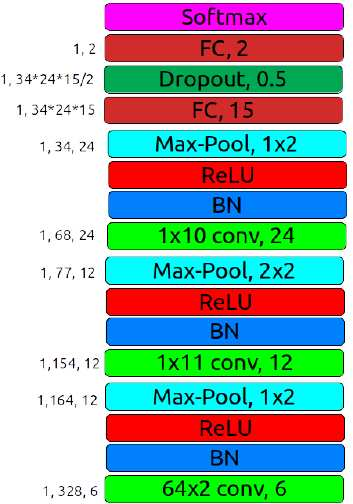}
    \caption{Deep ConvNet Architecture. Following EEG input, 3 number of convolutional layers one of which is 2-D and allows to preserve spatial relations. Output sizes (width, height and depth) corresponding to layers are given at the left side. Each pooling is performed without overlapping. BN and FC refers to batch normalization and fully-connected, receptively.}
    \label{fig1}
\end{figure}

In order to predict imagery tasks through EEG signals, we designed a deep ConvNet architecture in Figure \ref{fig1} inspired by successful study in \cite{ref24}. It consists of three convolution max-pooling layers, with first layer was dedicated to preserve spatial characteristics of EEG, followed by two traditional convolution layer, two fully-connected layers and a dropout layer (probability was set to 0.5). Batch normalization \cite{ref9} (\ref{eq1}) and rectified unit (ReLU) (\ref{eq2}) activation were applied following each discrete convolution operation at convolution layers.
\begin{equation}
\label{eq1}
    ReLU(x)=max(0,x)
\end{equation}
\begin{equation}
\label{eq2}
    H^{'}=\frac{H-\mu}{\sigma}
\end{equation}
where H is activation output of any layer to normalize, is a vector including means of each neuron and is a vector including standard deviation of each neuron. The training set was segmented in estimation and validation subsets (85 and 15\% of the training set respectively). Gradients were computed at each 100 batches, and weights were updated according to them with learning rate of 0.001 that decreases at level of 0.1 in each 10 epochs. 
Updating weights were separately realized with using three different approaches; stochastic gradient descend with momentum (SGDM) (momentum value was 0.9) optimization and adaptive moments (Adam) (gradient decay factor, squared gradient decay factor and epilson constant were 0.9, 0.99 and $10^{-8}$, respectively) and RmsPROP (squared gradient decay factor and epilson constant were 0.99 and $10^{-8}$, respectively) adaptive learning optimization. The training of the network was stopped either at the $100^{th}$ epoch or whenever the updates of the weights failed to reduce the loss (cross entropy) of the validation subset for 15 consecutive times. The status of the ConvNet was then reverted to the last most successful epoch. (Codes for downloading data form remote servers and guides for implementing this study in detail are available at \href{https://github.com/apdullahyayik/EEGMMI-Deep-ConvNET-}{https://github.com/apdullahyayik/EEGMMI-Deep-ConvNET-})

\begin{figure*}%
  \centering
  \subfigure[]{%
    \label{fig:A}%
    \includegraphics[width=0.4\textwidth]{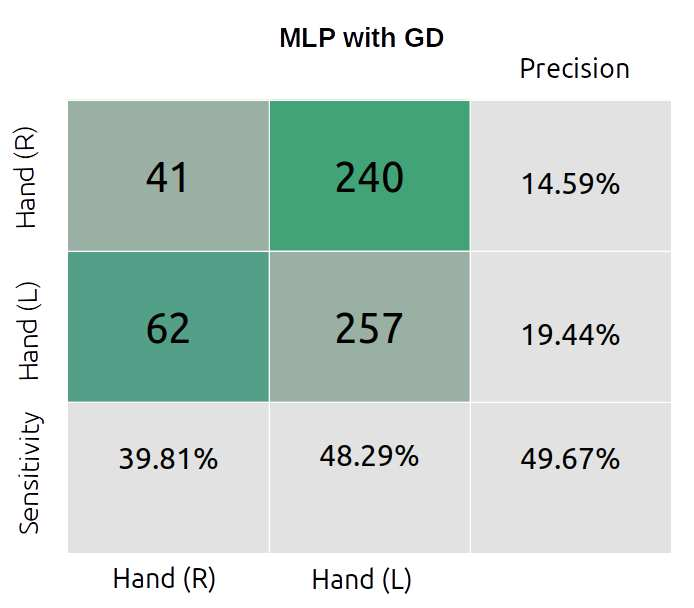}
    
            }%
  \subfigure[]{%
    \label{fig:B}%
   \includegraphics[width=0.4\textwidth]{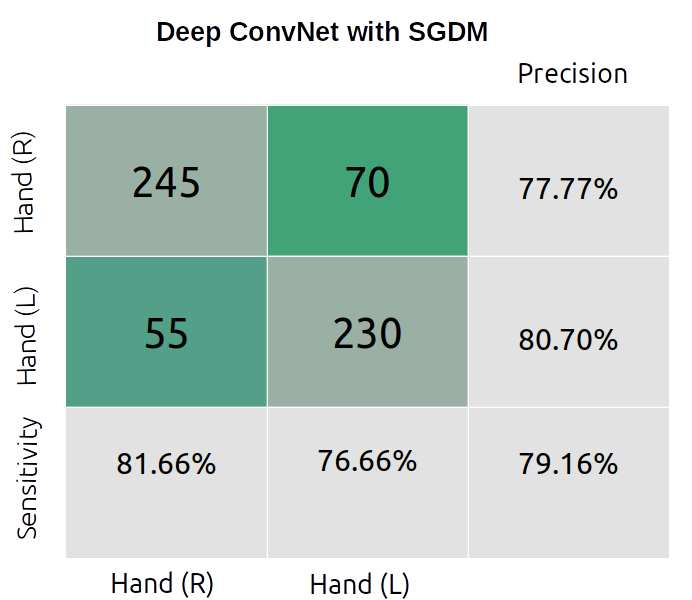}
            }%

  \subfigure[]{%
    \label{fig:C}%
    \includegraphics[width=0.4\textwidth]{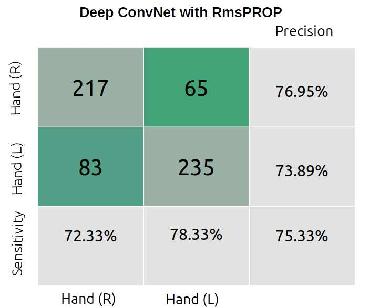}
            }%
  \subfigure[]{%
    \label{fig:D}%
    \includegraphics[width=0.4\textwidth]{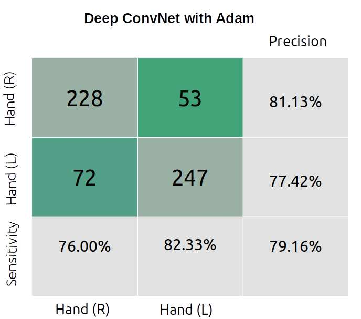}
           }

\caption{Confusion matrices for (a) Mutilayer perceptron (MLP) with gradient descent (GD), and for deep ConvNets with (b) stochastic gradient descent with momentum (SGDM), (c) RmsPROP and (d) adaptive momentum (Adam). Diagonal values correspond to accurately predicted numbers of trial for each class. Bottom rows correspond to sensitivity and right-most columns correspond to precision values. Lower-right values are overall accuracies.}
  \label{fig2}%
\end{figure*}

\subsection{Results and Conclusions}
Previous works in the literature predicted imagined hand movements on a subject-independent basis with considering only 20 number of subjects \cite{ref14}. 
In this study model, design and generalization capacity of the task were enhanced. We proposed a deep ConvNet approach for this challenging task through raw EEG data on a subject-independent basis with considering 109 number of subjects. For comparison handcrafted spectral features of Welch method were extracted and trained with multilayer perceptron. Confusion matrices and performance measures are detailed in Figure \ref{fig2}.

The fact that MLP with spectral features failed to predict, whereas deep ConvNet models achieved success indicates that the hierarchical feature representation and training strategies in deep ConvNets are suitable to modeling imagined motor movements on a subject dependent basis. In addition Adam and SGDM optimization techniques provided to reach accuracy of 79.16\% that is higher than RmsPROP. 

\section{Discussion}
\label{discussion}
This study shows that ConvNets allow accurate imagery hand movement predicting, that recent techniques; Adam optimization, batch normalization together with dropout strategy boost performance with raw EEG data, outperforming conventional fully-connected multilayer perceptron with hand-crafted spectral features.

Thus, ConvNets can provide robust learning from EEG data with only use of minimum preprocessing. This study also shows that ConvNets can offer promising achievements in neuroscience research field.

\section*{\small{Compliance with ethical standards}}
\textbf{\small{Conflict of interest}} \small{The authors declare that there is no conflict of interest.}

\bibliographystyle{spmpsci}      

\begin{thebibliography}{10}

\bibitem{ref1}
Yu-Te Wang, Yijun Wang, and Tzyy-Ping Jung.
\newblock A cell-phone-based brain--computer interface for communication in
  daily life.
\newblock {\em Journal of neural engineering}, 8(2):025018, 2011.

\bibitem{ref2}
Tom Carlson and Jose del~R Millan.
\newblock Brain-controlled wheelchairs: a robotic architecture.
\newblock {\em IEEE Robotics \& Automation Magazine}, 20(1):65--73, 2013.

\bibitem{ref3}
Lijuan Bai, Tianyou Yu, and Yuanqing Li.
\newblock A brain computer interface-based explorer.
\newblock {\em Journal of neuroscience methods}, 244:2--7, 2015.

\bibitem{ref4}
Nilufer Ozdemir and Esen Yildirim.
\newblock Patient specific seizure prediction system using hilbert spectrum and
  bayesian networks classifiers.
\newblock {\em Computational and mathematical methods in medicine}, 2014, 2014.

\bibitem{ref5}
Ian Goodfellow, Yoshua Bengio, and Aaron Courville.
\newblock {\em Deep learning}.
\newblock MIT press, 2016.

\bibitem{ref6}
Cicero Dos~Santos and Maira Gatti.
\newblock Deep convolutional neural networks for sentiment analysis of short
  texts.
\newblock In {\em Proceedings of COLING 2014, the 25th International Conference
  on Computational Linguistics: Technical Papers}, pages 69--78, 2014.

\bibitem{ref7}
Geoffrey Hinton, Li~Deng, Dong Yu, George Dahl, Abdel-rahman Mohamed, Navdeep
  Jaitly, Andrew Senior, Vincent Vanhoucke, Patrick Nguyen, Brian Kingsbury,
  et~al.
\newblock Deep neural networks for acoustic modeling in speech recognition.
\newblock {\em IEEE Signal processing magazine}, 29, 2012.

\bibitem{ref12}
Jason Sleight, Preeti Pillai, and Shiwali Mohan.
\newblock Classification of executed and imagined motor movement eeg signals.
\newblock {\em Ann Arbor: University of Michigan}, pages 1--10, 2009.

\bibitem{ref13}
MHF Zakaria, Wahidah Mansor, and Khuan~Y Lee.
\newblock Time-frequency analysis of executed and imagined motor movement eeg
  signals for neuro-based home appliance system.
\newblock In {\em TENCON 2017-2017 IEEE Region 10 Conference}, pages
  1657--1660. IEEE, 2017.

\bibitem{ref14}
Mohammad~H Alomari, Ali~M Baniyounes, and Emad~A Awada.
\newblock Eeg-based classification of imagined fists movements using machine
  learning and wavelet transform analysis.
\newblock {\em Int. J. Adv. Electron. Electr. Eng}, pages 83--87, 2014.

\bibitem{ref15}
Kalogiannis Gregory, Kapsimanis George, and Hassapis George.
\newblock An eeg pre-processing technique for the fast recognition of motor
  imagery movements.
\newblock In {\em 2016 IEEE Biomedical Circuits and Systems Conference
  (BioCAS)}, pages 90--94. IEEE, 2016.

\bibitem{ref16}
Hardik Meisheri, Nagraj Ramrao, and Suman Mitra.
\newblock Multiclass common spatial pattern for eeg based brain computer
  interface with adaptive learning classifier.
\newblock {\em arXiv preprint arXiv:1802.09046}, 2018.

\bibitem{ref17}
Dalin Zhang, Lina Yao, Xiang Zhang, Sen Wang, Weitong Chen, and Robert Boots.
\newblock Eeg-based intention recognition from spatio-temporal representations
  via cascade and parallel convolutional recurrent neural networks.
\newblock {\em arXiv preprint arXiv:1708.06578}, 2017.

\bibitem{ref8}
Nitish Srivastava, Geoffrey Hinton, Alex Krizhevsky, Ilya Sutskever, and Ruslan
  Salakhutdinov.
\newblock Dropout: a simple way to prevent neural networks from overfitting.
\newblock {\em The Journal of Machine Learning Research}, 15(1):1929--1958,
  2014.

\bibitem{ref9}
Sergey Ioffe and Christian Szegedy.
\newblock Batch normalization: Accelerating deep network training by reducing
  internal covariate shift.
\newblock {\em arXiv preprint arXiv:1502.03167}, 2015.

\bibitem{ref10}
Geoffrey Hinton, Nitish Srivastava, and Kevin Swersky.
\newblock Neural networks for machine learning lecture 6a overview of
  mini-batch gradient descent.
\newblock {\em Cited on}, 14, 2012.

\bibitem{ref11}
Diederik~P Kingma and Jimmy Ba.
\newblock Adam: A method for stochastic optimization.
\newblock {\em arXiv preprint arXiv:1412.6980}, 2014.

\bibitem{ref18}
Gerwin Schalk, Dennis~J McFarland, Thilo Hinterberger, Niels Birbaumer, and
  Jonathan~R Wolpaw.
\newblock Bci2000: a general-purpose brain-computer interface (bci) system.
\newblock {\em IEEE Transactions on biomedical engineering}, 51(6):1034--1043,
  2004.

\bibitem{ref19}
Ary~L Goldberger, Luis~AN Amaral, Leon Glass, Jeffrey~M Hausdorff, Plamen~Ch
  Ivanov, Roger~G Mark, Joseph~E Mietus, George~B Moody, Chung-Kang Peng, and
  H~Eugene Stanley.
\newblock Physiobank, physiotoolkit, and physionet: components of a new
  research resource for complex physiologic signals.
\newblock {\em Circulation}, 101(23):e215--e220, 2000.

\bibitem{ref20}
Peter Welch.
\newblock The use of fast fourier transform for the estimation of power
  spectra: a method based on time averaging over short, modified periodograms.
\newblock {\em IEEE Transactions on audio and electroacoustics}, 15(2):70--73,
  1967.

\bibitem{ref21}
Mike~X Cohen.
\newblock {\em Analyzing neural time series data: theory and practice}.
\newblock MIT press, 2014.

\bibitem{ref22}
Kunihiko Fukushima and Sei Miyake.
\newblock Neocognitron: A self-organizing neural network model for a mechanism
  of visual pattern recognition.
\newblock In {\em Competition and cooperation in neural nets}, pages 267--285.
  Springer, 1982.

\bibitem{ref23}
Yann LeCun, L{\'e}on Bottou, Yoshua Bengio, Patrick Haffner, et~al.
\newblock Gradient-based learning applied to document recognition.
\newblock {\em Proceedings of the IEEE}, 86(11):2278--2324, 1998.

\bibitem{ref24}
Robin~Tibor Schirrmeister, Jost~Tobias Springenberg, Lukas Dominique~Josef
  Fiederer, Martin Glasstetter, Katharina Eggensperger, Michael Tangermann,
  Frank Hutter, Wolfram Burgard, and Tonio Ball.
\newblock Deep learning with convolutional neural networks for eeg decoding and
  visualization.
\newblock {\em Human Brain Mapping}, aug 2017.

\end{thebibliography}

\end{document}